\documentclass[twocolumn, amssymb, amsmath, aip, apl, showpacs, 10pt]{revtex4-1}
\usepackage[dvips]{graphicx}
\usepackage[dvips]{color}
\usepackage{hyperref}
\hypersetup{
    colorlinks=true,
    linkcolor=blue,
}

\begin{document}
\title{Multifunctional behavior of Mn-site doped antiferromagnetic Mn$_5$Si$_3$ alloys}
\author{S. C. Das}
\author{S. Pramanick}
\author{S. Chatterjee}
\email{souvik@alpha.iuc.res.in}
\affiliation{UGC-DAE Consortium for Scientific Research, Kolkata Centre, Sector III, LB-8, Salt Lake, Kolkata 700 106, India}
\begin{abstract}
Present work reports a detailed investigation on the magnetoresistance and magnetocaloric behavior of Ni and Cr-doped Mn$_5$Si$_3$ alloys with general formula Mn$_{5-x}$A$_x$Si$_3$ (where A = Ni/Cr; $x$ = 0, 0.05, 0.1 and 0.2). Both pure (undoped) and doped alloys show a reasonably large amount of magnetoresistance (MR). Doping at Mn-site, both by Ni and Cr, results in a monotonic decrease in MR values. Magnetocaloric effect (MCE), on the other hand, is found to be interesting, and all the alloys show both conventional and inverse MCE around the magneto-structural transition temperature. Among the two types of MCE observed, the inverse MCE is found to decrease with increasing doping concentration and consistent with the MR behavior, whereas doping results in a significant increase in conventional MCE values.
\end{abstract}
\maketitle

The recent discovery of different interesting physical and functional properties, like, magnetocaloric effect, anomalous Hall effect, inverted hysteresis behavior, thermomagnetic irreversibility and spin fluctuation in a manganese-silicide alloy of nominal composition Mn$_5$Si$_3$ triggered renewed interest about the material among researchers in last one decade~\cite{lander,menshikov,silva,gottschilch,brown,kanani,surgers,biniskos,prb-scd,scd-unpublished}. At room temperature, Mn$_5$Si$_3$ alloy is found to be paramagnetic (PM) in nature and adopts hexagonal $D8_8$ structure with space group $P6_3/mcm$~\cite{lander,menshikov}. Below 100 K, it becomes orthorhombic (space group $Ccmm$) and ordered in collinear antiferromagnetic (AFM2) fashion, whereas, non-collinear antiferromagnetic (AFM1) ordering with monoclinic structure (non-centrosymmetric $Cc2m$ space group) has been observed below 66 K~\cite{lander,menshikov}. Both these transitions are first order in nature. In addition to these two antiferromagnetic phases (AFM1 and AFM2), S\"urgers {\it et al.} have recently reported the presence of another non-collinear antiferromagnetic phase (AFM1$^{\prime}$) which appears only in the presence of external magnetic field ($H$)~\cite{surgers}. Our recent work confirmed the presence of such AFM1$^{\prime}$ phase and established its crucial role behind the observation of unusual properties like inverted hysteresis behavior and thermomagnetic irreversibility~\cite{prb-scd,scd-unpublished}. 

\par
Alloys/compounds having first-order structural transition found to be the right candidate for the magneto-functional applications. Though several works have been performed on the pure and doped Mn$_5$Si$_3$ alloys to address different physical properties, the magneto-functional properties of these alloys have yet not been explored. In this particular work, we shall discuss some of the magneto-functional features, which include magnetoresistance (MR) and magnetocaloric effect (MCE), of some Ni and Cr-doped Mn$_5$Si$_3$ alloys. Recently, we have successfully prepared Ni and Cr-doped Mn$_5$Si$_3$ alloys to check the robustness of the inverted hysteresis behavior and thermomagnetic irreversibility observed in pristine Mn$_5$Si$_3$ alloy~\cite{scd-unpublished}. Doping in Mn-site significantly affects these unusual properties~\cite{scd-unpublished}. Observation of such a strong effect of Mn-site doping on the magnetic properties and presence of first-order phase transition influenced us to investigate the unexplored MR and MCE properties of doped Mn$_5$Si$_3$ alloys. MR of a material is defined as the change in electrical resistance in the presence of an external magnetic field ($H$), and it is important for memory application. MCE, on the other hand, is associated with the change in temperature of magnetic solids in a changing magnetic field~\cite{gs-mce,gs-prl1}. Adiabatic change in entropy ($\Delta S$) is a measure of MCE, and it plays a vital role in the field of environment-friendly magnetic cooling technology. This work is probably the first-ever approach to explore such properties in Mn-site doped Mn$_5$Si$_3$ alloys. For the sake of completeness, we have compared the results of the doped alloys with pure Mn$_5$Si$_3$ alloys.

\begin{figure*}[t]
\centering
\includegraphics[width = 17 cm]{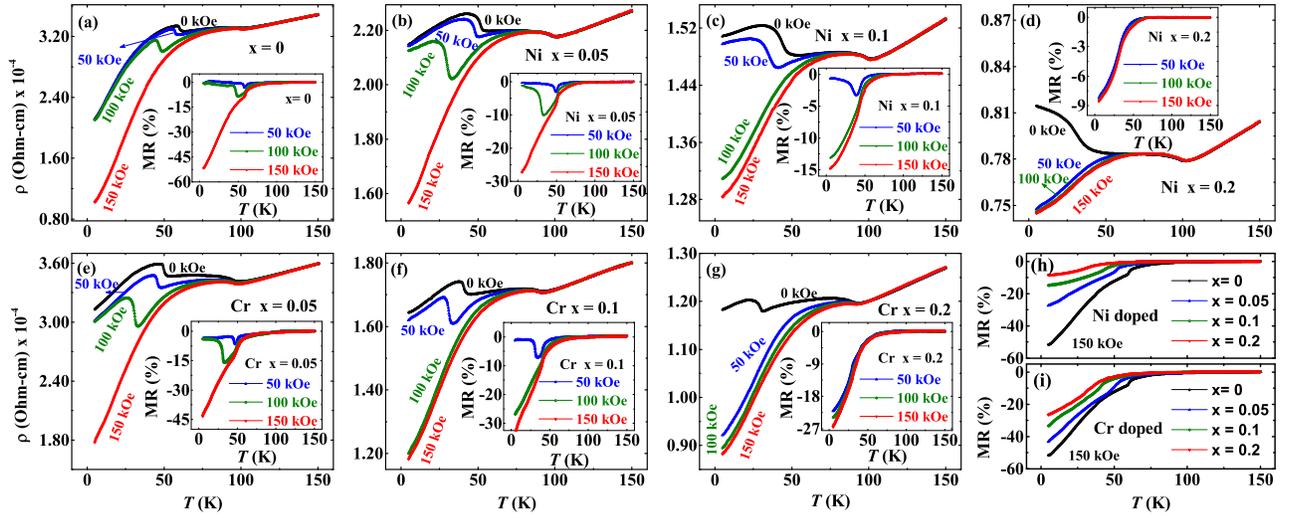}
\caption{(Color online) Temperature ($T$) variation of dc electrical resistivity ($\rho$) in presence of 0, 50, 100 and 150 kOe of external magnetic field ($H$) are plotted in the main panels for pure ((a) $x$ = 0), Ni-doped ((b) $x$ = 0.05, (c) $x$ = 0.10, (d) $x$ = 0.20) and Cr-doped ((e) $x$ = 0.05, (f) $x$ = 0.10, (g) $x$ = 0.20) alloys. Insets of (a)-(g) depict $T$ variation of magentoresistance (MR) of corresponding alloys in presence of 50, 100 and 150 kOe of external $H$. $T$ variation of MR for Ni and Cr-doped alloys along with the undoped alloy in presence of 150 kOe of external $H$ are plotted in (h) and (i) respectively.}
\label{mr1}
\end{figure*}

\par
The polycrystalline alloys of nominal compositions Mn$_{5-x}$A$_x$Si$_3$ (where A = Ni and Cr; $x$ = 0, 0.05, 0.1, and 0.2) were prepared by the arc-melting method as mentioned in our previous works~\cite{prb-scd,scd-unpublished}. The samples were characterized using x-ray powder diffraction technique. All the samples are found to show D8$_8$ type hexagonal structure, and no impurity phase has been detected in any of the prepared alloys. Dc magnetization ($M$) and electrical resistivity ($\rho$) of the samples were measured using a commercial cryogen-free 150 kOe system from Cryogenic Ltd., UK. During measurements, helium exchange gas was used for better temperature stability. At the time of isothermal measurements, the observed temperature fluctuation was found to be less than 15 mK. The standard four-probe method was adopted for $\rho$ measurements, and $H$ was applied perpendicular to the direction of the current. 
\par
The electrical resistivity of the pure and doped (both Ni and Cr-doped) alloys, recorded as a function of temperature ($T$) in the presence of different applied $H$, is depicted in the main panels of fig.~\ref{mr1} (a-g). The $\rho$ of the presently studied samples was recorded in the cooling protocol. The zero-field $\rho(T)$ data for pure and doped alloys show clear anomalies around the magnetic transitions present in these alloys [AFM1 to AFM2 ($T_{N1}$) and AFM2 to PM ($T_{N2}$)]. A monotonic decrease in $T_{N1}$ has been observed with increasing doping concentration. On the other hand, $T_{N2}$ remains almost unchanged with doping. The value of the various transition temperatures observed from the $\rho(T)$ data is consistent with our previous works on the pure and doped alloys~\cite{prb-scd,scd-unpublished}. Application of external $H$ results in a significant decrease in $T_{N1}$, and with high enough $H$ (value of the critical field depends strongly on the sample composition), such transition vanishes, and the alloys remain in collinear AFM2 phase till the lowest temperature of measurements. A similar effect has also been observed in the case of the magnetization data of the pure and doped alloys~\cite{prb-scd,scd-unpublished}. The $\rho(T)$ curve in the presence of external $H$ starts to deviate from the zero-field data below/around $T_{N2}$, and such deviation becomes more prominent below/around $T_{N1}$. We have calculated and plotted MR $\left[\rm{MR}(\%)=\frac{\rho(H)-\rho(0)}{\rho(0)}\times 100 (\%)\right]$ as a function of $T$ at three different constant $H$ (= 50 kOe, 100 kOe and 150 kOe) in the insets of fig.~\ref{mr1} (a-g). In the low $H$ region (till the existence of AFM1 to AFM2 transition), MR is found to be maximum around the $T_{N1}$ for $x$ = 0, 0.05, and 0.1 alloys (both Ni and Cr-doping cases). But in the presence of 150 kOe of applied $H$ (well above AFM1 to AFM2 critical transition fields), the maximum value of MR has been observed at the lowest $T$ of measurements for these alloys. On the other hand, for both 4\% Ni and Cr doped alloys ($x$ = 0.2), 50 kOe $H$ is high enough to make AFM1 to AFM2 transition disappear from the alloys, and maximum value of MR has been observed at 5 K. MR($T$) data recorded in the presence of 10 kOe of external $H$ for the 4\% doped alloys show similar behavior as that of the pure, 1\%, and 2\% doped alloys (not shown here). Notably, the magnitude of the MR observed for pure and doped alloys are reasonably high. For the sake of completeness, we plotted MR {\it vs.} $T$ data in the presence of 150 kOe of external $H$ for both Ni and Cr doped alloys along with the pure Mn$_5$Si$_3$ alloy (see fig.~\ref{mr1} (h) and (i)). From the MR {\it vs.} $T$ data, it is clear that the MR is found to be maximum for the undoped Mn$_5$Si$_3$ alloy and a monotonic decrease in MR values has been observed with increasing Ni/Cr concentration. At 5 K, 53\% MR has been observed in the presence of 150 kOe of external $H$ for the undoped Mn$_5$Si$_3$ alloy. Whereas, 8\% and 25\% MR have been observed for 4\% Ni and Cr doped alloys ($x$ = 0.2) respectively at 5 K for $H$ = 150 kOe.

\par
\begin{figure}[t]
\centering
\includegraphics[width = 8 cm]{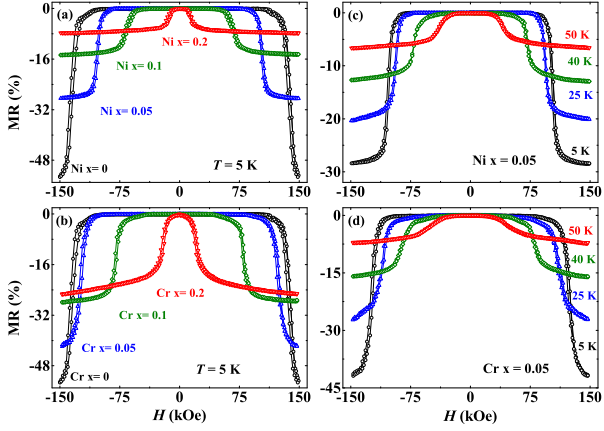}
\caption{(Color online) (a) \& (b) represents magnetoresistance (MR) as a function of the magnetic field ($H$) at 5 K for Ni and Cr-doped alloys along with the undoped alloys, respectively. Isothermal MR {\it vs.} $H$ data 1\% Ni and Cr-doped alloys ($x$ = 0.05) at different constant temperatures are plotted in (c) and (d) respectively.}
\label{mrh}
\end{figure}

To shed more light on the MR properties of the presently studied alloys in detail, we recorded isothermal $\rho$($H$) data at different constant temperatures in the zero-field cooled protocol. The MR calculated from the 5 K isotherms for Ni and Cr-doped alloys along with the pristine Mn$_5$Si$_3$ alloy are plotted as a function of external $H$ in fig.~\ref{mrh} (a) and (b) respectively. Signature of AFM1 to AFM2 field-induced transition is visible in all the studied alloys. A notable decrease in the critical field for such AFM1 to AFM2 field-induced transition with increasing doping concentration has been observed. Interestingly, no effect of field-induced AFM1 to AFM1$^{\prime}$ transition has been observed in the isothermal $\rho(H)$ (hence in MR($H$)) data for any of the studied alloys. This is unlike the isothermal $M(H)$ data of the pure and doped alloys, where a clear signature of field-induced AFM1 to AFM1$^{\prime}$ transition is visible~\cite{scd-unpublished}. The non-collinear nature of both AFM1 and AFM1$^{\prime}$ phases that offers a similar kind of scattering for the electrons, resulting in similar values of $\rho$ in both AFM1 and AFM1$^{\prime}$ phases, may play a pivotal role behind such observation. We also observed a significant change like the MR($H$) curve with an increasing sample $T$. MR {\it vs.} $H$ at different constant $T$ for one representative sample from each of Ni and Cr-doped family ($x$ = 0.05) are plotted in fig.~\ref{mrh} (c) \& (d) respectively. The magnitude of MR and the critical field for AFM1 to AFM2 transition, both are found to decrease monotonically with increasing sample $T$ for all Ni and Cr-doping cases. Such behavior is consistent with the MR {\it vs.} $T$ and isothermal $M(H)$ data.

\par
\begin{figure}[t]
\centering
\includegraphics[width = 8.0 cm]{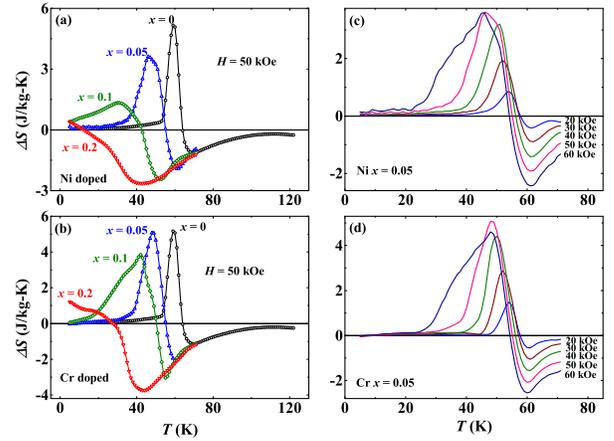}
\caption{(Color online) Field-induced entropy change ($\Delta S$) as a function of $T$ for the applied field ($H$) changing from 0$\rightarrow$50 kOe for Ni and Cr-doped alloys along with the undoped alloy are depicted in (a) \& (b) respectively. (c) and (d) show $\Delta S$ {\it vs.} $T$ data recorded for different values of changing $H$ for 1\% Ni and Cr-doped alloys ($x$ = 0.05) respectively.}
\label{mce}
\end{figure}

Observation of notable amount of MR and large change of dc $M$ values around $T_{N1}$ in the presently studied pure and doped alloys are the indications of the possible appearance of a significant amount of MCE. Keeping such a possibility in mind, MCE of all the doped alloys along with the pure Mn$_5$Si$_3$ alloy have been investigated. Using Maxwell's thermodynamical relation $\Delta S(0\rightarrow H_0)$ = $\int_0^{H_0}\left(\frac{\partial M}{\partial T}\right)_HdH$, the adiabatic change in entropy ($\Delta S$), which is a measure of MCE, can easily be calculated from the dc magnetization data~\cite{gs-mce}. Discontinuous jump in $M(H)$ or $M(T)$ data around the first-order phase transition sometimes gives rise to an infinite value of $\left(\frac{\partial M}{\partial T}\right)_H$ which may result in an erratic value of $\Delta S$~\cite{Oliveira-prb}. Absence of any kind of discontinuity in $M(H)$ and $M(T)$ data of the presently studied alloys allowed us to use Maxwell's thermodynamical relation for $\Delta S$ calculation. We recorded isothermal $M(H)$ data for all the studied alloys with 3 K intervals (between 5-122 K for the pure alloy and between 5-71 K for the doped alloys; not shown here). Such $M(H)$ isotherms were convoluted to obtain the isofield $M(T)$ curves that have been utilized to calculate $\Delta S$ using Maxwell's relation. Variation of $\Delta S$ with $T$ for external $H$ changing from 0$\rightarrow$50 kOe of Ni and Cr-doped alloys along with the pure Mn$_5$Si$_3$ alloy are depicted in fig.~\ref{mce} (a) and (b) respectively. Interestingly, except $x$ = 0.2 alloys, pure and all other doped alloys show a clear signature of the inverse (positive) and conventional (negative) MCE. On the other hand, the signature of inverse MCE is not clear for $x$ = 0.2 alloys. Lower values of AFM1 to AFM2 transition temperature in $x$ = 0.2 alloys are playing the key role behind such observations. Both the inverse and conventional peak temperatures are found to be shifted towards the lower temperature with increasing doping concentration. Interestingly, the magnitude of the inverse and the conventional MCEs behave differently. The undoped Mn$_5$Si$_3$ alloy shows maximum inverse MCE with a peak value of 5.2 J/kg-K for $H$, changing from 0$\rightarrow$50 kOe. With increasing doping concentration, the peak value of $\Delta S$ decreases, and for 2\% Ni and Cr-doped alloys ($x$ = 0.1), the peak magnitude becomes 1.3 and 3.9 J/kg-K respectively. On the other hand, a significant increase in conventional MCE behavior has been observed. The peak values of pure, 4\% Ni-doped, and 4\% Cr-doped alloys are found to be -1.1, -2.6, and -3.7 J/kg-K respectively for $H$ changing from 0$\rightarrow$50 kOe. Besides, the peak shape of the $\Delta S(T)$ curve around the inverse, and conventional MCE regions are found to be getting broader with increasing doping concentration. 

\par
Refrigeration capacity (RC) is another important parameter that we have calculated from the $\Delta S(T)$ data for the presently studied alloys around the inverse MCE region. RC is defined as RC = -$\int_{T_{\rm cold}}^{T_{\rm hot}} \Delta SdT$, where $T_{\rm hot}$ and $T_{\rm cold}$ are source and sink temperatures respectively. The calculated value of RC for the pure alloy is about -31.2 J/kg. RC values behave non-monotonically with Ni-doping and monotonically with Cr-doping. We observe a steady increase in RC with Cr-doping. For 1\% and 2\% Cr-doped alloys, the RC values are found to be -51.0 and -78.2 J/kg. On the other hand, RC values 1\% and 2\% Ni-doped alloys are -46.8 and -31.2 J/kg. We have also checked the MCE behavior of the studied alloys for different values of applied $H$. $\Delta S(T)$ data for two representative alloys (1\% Ni and Cr-doped alloys) are plotted in fig.~\ref{mce} (c) and (d). Both inverse and conventional MCE show a monotonic increase with increasing field magnitude. The peak positions of $\Delta S(T)$ data behave differently with external changing $H$. We observe a steady decrease in peak temperature around the inverse MCE region with increasing $H$, whereas the peak temperature remains unchanged around the conventional MCE region.

The present work deals with a systematic investigation of the MR and MCE properties of pure, Ni, and Cr-doped Mn$_5$Si$_3$ alloys of nominal compositions Mn$_{5-x}$A$_x$Si$_3$ (where A = Ni/Cr; $x$ = 0, 0.05, 0.1, and 0.2). All the studied alloys show clear signatures of PM-AFM2-AFM1 transitions during cooling from room temperature (300 K) in the $\rho(T)$ data. Both Ni and Cr-doping results in a significant decrease in the $T_{N1}$, whereas $T_{N2}$ remains unchanged. Application of external $H$ shows a similar effect on the transition temperatures as that of the dc magnetic data of the Mn-site doped alloys~\cite{prb-scd}. The large differences in dc $\rho$ values between AFM1 and AFM2 phases result in a significantly large MR in the presently studied alloys as the application of $H$ prefers the AFM2 phase. Around 53\% MR in the presence of 150 kOe of external $H$ at 5 K has been observed for the undoped Mn$_5$Si$_3$ alloy. Mn-site doping (both by Ni and Cr) affects the MR properties, and a significant decrease in MR magnitude has been observed. 4\% Ni and Cr doped alloys show 8\% and 25\% of MR respectively at 5 K in the presence 150 kOe of external $H$. All the alloys show a clear signature of field-induced AFM1-AFM2 transition in isothermal $\rho(H)$ (and hence MR($H$)) data. No signature of AFM1-AFM1$^{\prime}$ transition in MR($H$) data has been observed in any of the alloys, unlike the $M(H)$ isotherms, where such field-induced transitions are visible~\cite{surgers,biniskos,prb-scd,scd-unpublished}. Apart from MR, the pure and doped alloys show a reasonably large value of MCE. Both inverse and conventional MCE are visible in the presently studied alloys. The peak values of the $\Delta S(T)$ curve around the inverse, and conventional MCE regions found to behave differently with doping concentration. The most important observation is the increment of conventional MCE values with increasing doping concentration. 4\% of Ni and Cr doping in the Mn site of the Mn$_5$Si$_3$ alloy results in a 180\% and 290\% increase in peak values of the $\Delta S(T)$ curve (for $H$ changing from 0$\rightarrow$50 kOe) around the conventional MCE region respectively. Observation of large values of both MR and MCE in the presently studied alloys makes them a potential candidate for practical applications. These doped alloys are an excellent addition to the existing members of the Mn-Si family of alloys.

SCD (IF160587) would like to thank DST, India, for providing Inspire fellowship.

%

\end{document}